\documentclass{article}

\usepackage{arxiv}

\usepackage[utf8]{inputenc} 
\usepackage[T1]{fontenc}    
\usepackage{hyperref}       
\usepackage{url}            
\usepackage{booktabs}       
\usepackage{amsfonts}       
\usepackage{nicefrac}       
\usepackage{microtype}      
\usepackage{lipsum}		
\usepackage{graphicx}
\usepackage{natbib}
\usepackage{doi}

\title{On the Challenges and Perspectives of Foundation Models for Medical Image Analysis}

\author{{Shaoting Zhang}
  \\
	Shanghai AI Laboratory\\
	Shanghai, China\\
 \texttt{zhangshaoting@pjlab.org.cn}\\
	\And
	{Dimitris Metaxas} \\
	Rutgers University\\
	New Brunswick, NJ, USA\\
	\texttt{dnm@cs.rutgers.edu} \\
}

\renewcommand{\shorttitle}{\textit{arXiv} Template}

\hypersetup{
pdftitle={A template for the arxiv style},
pdfsubject={q-bio.NC, q-bio.QM},
pdfauthor={David S.~Hippocampus, Elias D.~Striatum},
pdfkeywords={First keyword, Second keyword, More},
}

\begin{document}
\maketitle

\begin{abstract}
This article discusses the opportunities, applications and future directions of large-scale pretrained models, i.e., foundation models, which promise to significantly improve the analysis of medical images. 
Medical foundation models have immense potential in solving a wide range of downstream tasks, as they can help to accelerate the development of accurate and robust models, reduce the dependence on large amounts of labeled data, preserve the privacy and confidentiality of patient data. 
Specifically, we illustrate the ``spectrum'' of medical foundation models, ranging from general imaging models, modality-specific models, to organ/task-specific models, and highlight their challenges, opportunities and applications. We also discuss how foundation models can be leveraged in downstream medical tasks to enhance the accuracy and efficiency of medical image analysis, leading to more precise diagnosis and treatment decisions.

\end{abstract}


\section{Introduction}
\label{sec1}

\begin{figure*}[h]
  \centering
  \includegraphics[width=0.75\textwidth]{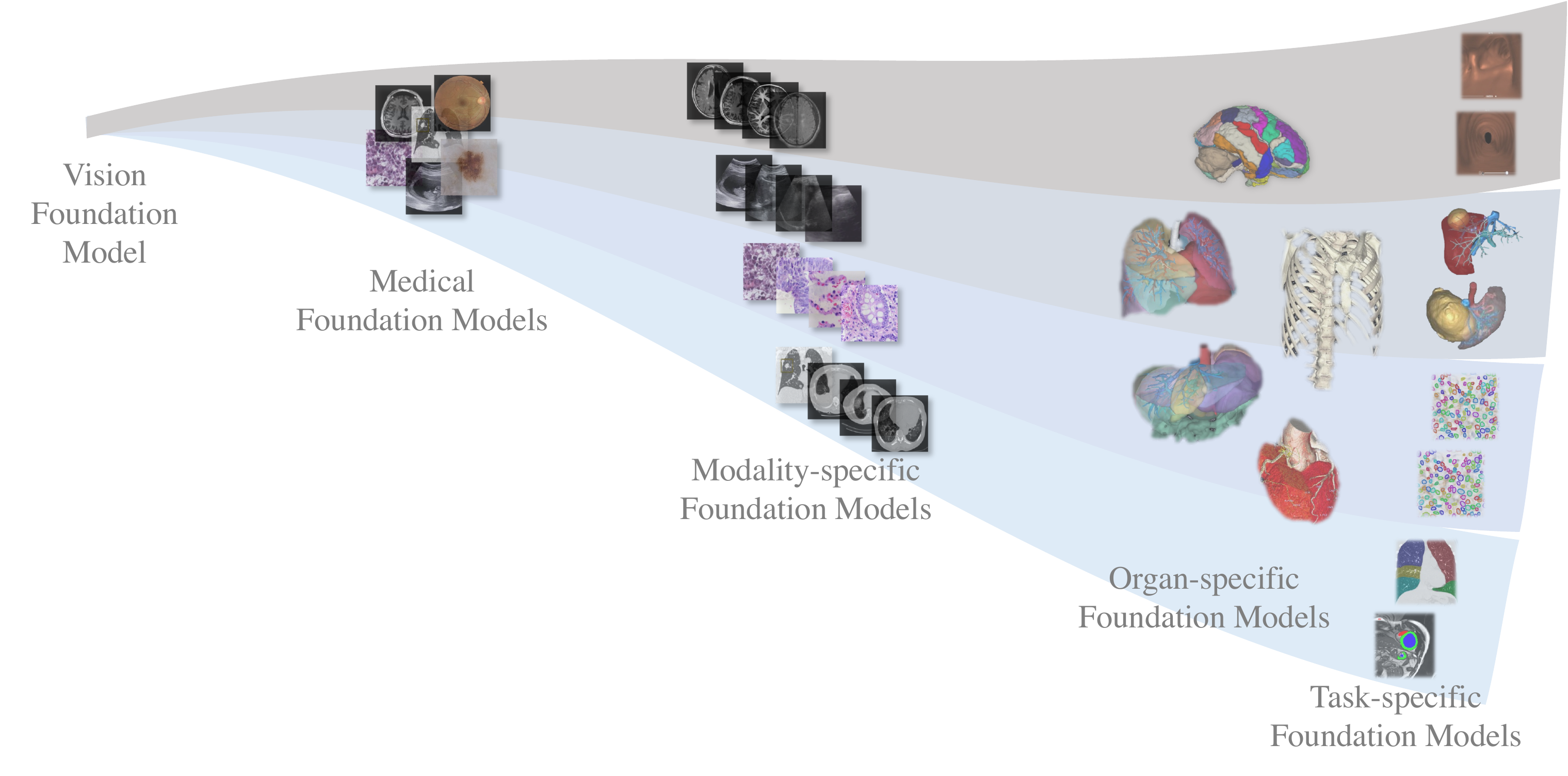}
  \vspace{-1em}
  \caption{The spectrum of foundation models in medical image analysis.}
  \label{fig:spectrum}
\end{figure*}

A salient distinction exists between traditional pretrained models and contemporary foundation models. The former~\cite{deng2009imagenet,vaswani2017attention,devlin2018bert,radford2018improving,radford2019language,dosovitskiy2021an} typically require extensive supervised fine-tuning to address specific downstream tasks, whereas the latter are capable of employing few-shot learning, zero-shot learning, or prompt engineering to manage a wide variety of tasks with a singular set of model weights~\cite{brown2020language,chowdhery2022palm,ouyang2022training,touvron2023llama,touvron2023llama2,google2023palm2}. Consequently, foundation models exhibit considerable generalizability and adaptability, positioning them as a focal point of recent machine learning (ML) research.
 
In the field of medical image analysis, however, task-specific ML models are still the main methods used, especially for clinical applications such as computer-aided disease diagnosis. Developing medical foundation models presents a significant challenge due to the diverse imaging modalities used in medicine, which differ greatly from natural images and are based on a spectrum of physics-based properties and energy sources. These modalities are based on the use of light, electrons, lasers, X-rays, ultrasound, nuclear physics, and magnetic resonance. The images produced span multiple scales, ranging from molecules and cells to organ systems and the full body. Therefore, it may be infeasible to develop a unified multi-scale foundation model trained from a combination of these multi-modality images. 

In the following, we will investigate and present our vision for the ``spectrum'' of foundation models and their uses in medical image analysis, ranging from general vision models, modality-specific models, to organ and task-specific models (Fig.~\ref{fig:spectrum}). Fortunately, the growing availability of high-quality publicly available annotated medical data has led to the gradual emergence of specialized foundational models with an innate capacity for generating more generalized representations of medical data. Therefore, foundation models trained with medical images and/or natural images in a self-supervised manner may serve as an improved solution basis for important clinical problems, will result in advances in the field of medical imaging, and will improve the efficacy and efficiency of disease diagnosis and treatment.

\section{The Spectrum of Foundation Models}

\noindent\textbf{Vision Foundation Models}

A straightforward approach is to employ foundation models trained from natural images~\cite{carion2020end,dosovitskiy2021an,zhai2022scaling,radford2021learning,he2022masked,wang2022internimage,oquab2023dinov2}, and then design sophisticated algorithms to solve downstream medical tasks. 

However, the lack of publicly available quality annotations in medical imaging has been the bottleneck for training large-scale deep learning models for many downstream clinical applications. It remains a tedious and time-consuming job for medical professionals to hand-label image data repeatedly, while providing a few differentiable sample cases is feasible and complies with the training process of medical residents. Vision foundation models, often trained on large-scale visual images of various modalities, could serve as the basis for building medical applications.

However, while vision foundation models can learn general representations, medical images have unique characteristics whose features and patterns differ significantly from those in natural images. Therefore, carefully designed algorithms are necessary to adapt them to domain-specific medical problems. Fine-tuning, additional adapters, prompting strategies, and specialized architectures or modifications are potential solutions to achieve optimal performance in medical problems.

For instance, the recent Segmentation Anything Model (SAM)~\cite{kirillov2023segany}, a promptable segmentation system with zero-shot generalization to unfamiliar objects and images, has demonstrated its impressive performance on natural images. 
However, its out-of-the-box performance on complex medical tasks such as pancreas, spine or cell nuclei segmentation is not satisfactory~\cite{he2023accuracy,roy2023sam,deng2023segment,mazurowski2023segment,shi2023generalist}.
SAM can be further tuned to achieve state-of-the-art performance by leveraging high-quality downstream data and performing proper fine-tuning strategies~\cite{ma2023segment,paranjape2023adaptivesam,cui2023all}, adding adapters with specially designed architectures~\cite{wu2023medical,gong20233dsam,chen2023ma}, or effective prompts~\cite{huang2023segment,cheng2023sam} with manual annotations. Venturing a step further, we could explore the possibility of combining the output of localization/detection algorithms with SAM or integrating SAM with image processing and visualization software like 3D slicer~\cite{liu2023samm}. This fusion would pave the way for a robust pipeline tailored for complex medical applications.

In general, a unified foundation model approach can not achieve state-of-the-art performance in many medical image analysis tasks due to large variations present in organs and important structures, texture, shape, size and topology (e.g., blood vessels), and imaging modalities.  
Furthermore, it is noteworthy to mention that there are parameter/data efficient tuning methods to adapt vision foundation models to address image analysis challenges arising from long-tail medical data.

\noindent\textbf{Modality-specific Foundation Models}

Depending on the pathology, various types of imaging modalities are employed for 
diagnostic and therapeutic purposes. They include X-rays, Computed Tomography (CT), Magnetic Resonance Image (MRI), Ultrasound imaging, and Positron Emission Tomography (PET).
A modality-specific foundation model is specifically designed for a group of imaging modalities such as radiology images (including X-ray, CT, MR, and Ultrasound)~\cite{ghesu2022self}, 3D images (stacks of 2D CT and MR images)~\cite{chen2019med3d}, or a particular medical imaging modality which includes X-ray~\cite{tiu2022expert}, CT~\cite{huang2023stu,wang2023mis}, endoscopy~\cite{wang2023foundation}, and pathology~\cite{chen2022scaling,chen2023general,vorontsov2023virchow} images.
It is then used to learn image-based features that are relevant to the intended use of the particular modality. For example, a CT-specific model may learn to identify features related to bone density and tissue contrast, while an MRI-specific model may learn to identify features related to soft tissue contrast and motion.

Vision foundation models trained on large-scale natural image datasets, can provide a strong starting point for a wide range of medical imaging analysis tasks. Using these vision foundation models, modality-specific foundation models can leverage the unique characteristics of each imaging modality and can result in models optimized for specific modalities. While they can then lead to higher accuracy and efficiency for the analysis tasks specific to that modality, they may not generalize well to other modalities.

\noindent\textbf{Organ/Task-specific Foundation Models}

\begin{figure}[h]
  \centering
  \includegraphics[width=0.65\linewidth]{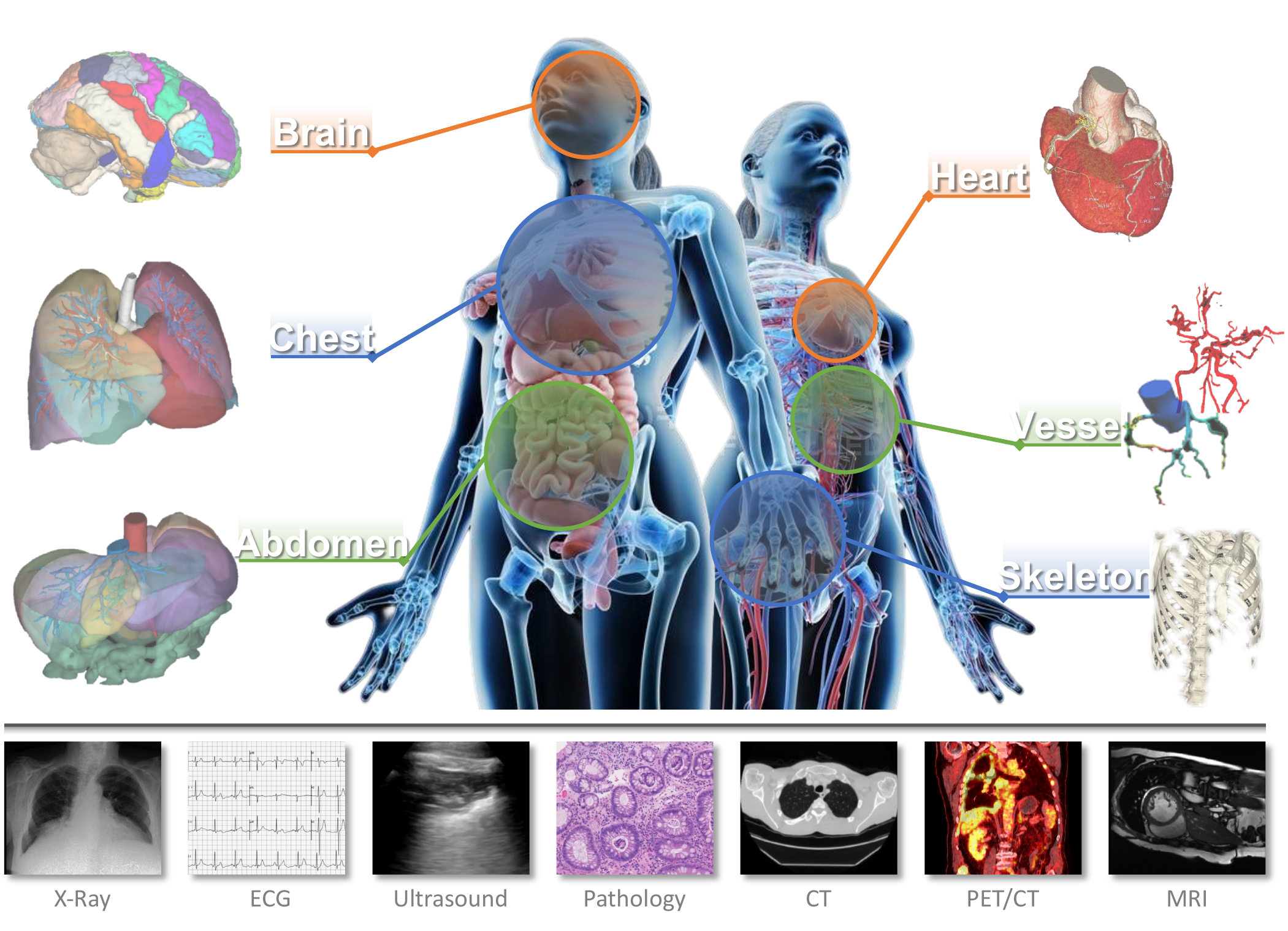}
  \caption{Different image modalities have large image-level variations, which may cause difficulties when training a unified foundation model. Similarly each modality given the imaging formation differences, results in images with significant variations in organ appearance and related structures which determine which modality needs to be used given a target organ pathology.
  Given modalities, leveraging fine-grained models for learning organ appearance and pathology, will enable important clinical methods and tools such as robust computer-aided diagnosis and surgical planning.}
  \label{fig:organ}
\end{figure}

More specifically, the foundation models could be tailored to a particular medical organ~\cite{li2020self,luo2022word,zhou2023foundation} or diagnostic task, such as segmentation~\cite{antonelli2022medical,tang2022self,butoi2023universeg}. This use aims to address the challenges posed by the variability in organ appearance across medical images, as well as the diverse range of clinical tasks that are based on image analysis (Fig.~\ref{fig:organ}). 

Collecting data for training organ/task-specific foundation models can be challenging due to the need for large amounts of labeled data. Nevertheless, a well-trained organ/task-specific foundation model can provide better accuracy and interpretability as well as significantly reduce the amount of labeled data required for new tasks, as it has already learned relevant features from the previous training.

\noindent\textbf{General vs. Specialized Foundation Models}

By definition, specialists possess an in-depth understanding of a particular subject matter, whereas generalists have a broader purview, either within a single field or across multiple disciplines.

In the context of medical image analysis, a general AI system is characterized as a multitask and multimodal platform capable of performing a diverse range of tasks on multimodal images of different organs and diseases. They include classification, detection, segmentation and registration and utilize a single set of model weights~\cite{moor2023foundation,tu2023towards,wu2023towards}. This approach and related models are shown in the left part of Fig.1, which shows the spectrum of foundation models. 

Conversely, specialized AI systems are designed to perform discrete clinical tasks, such as the detection of pulmonary nodules, the reconstruction of coronary arteries, or the diagnosis of hepatocellular carcinoma. These systems are generally confined to a particular organ and imaging modality, aligning more closely with the right side of the foundation model spectrum (see Fig.1).

Within the computer science community, there is an emerging focus on the development of general AI frameworks. This shift is driven by the technical innovation inherent in exploring large, multimodal generative models capable of processing diverse types of medical data. However, research in both academia, medical institutions, and industry, largely remains focused on the development of specialized AI systems. 
This focus is attributable to several factors. Firstly, most existing state-of-the-art medical image analysis systems use a single type of imaging modality (unimodal) and are trained on a single task such as segmentation or classification. Secondly, AI currently serves primarily as an assistant to medical professionals who require targeted support, which is consistent with their medical training. Consequently, there is a pragmatic inclination toward designing specialized systems that excel in terms of performance and accuracy in particular tasks. Moreover, general type of AI systems tend to consume significantly greater computational resources and often lack the required accuracy.

Both specialized and general AI system approaches offer distinct advantages and are suited for different applications. Accordingly, we advocate for a comprehensive exploration of the foundation model spectrum to ascertain the optimal trade-off between developmental effort and practical efficacy.

\section{Data Requirements for Foundation Models}

\noindent\textbf{Data to Pretrain the Foundation Models}

Data is the cornerstone for training all foundation models. Preparing data for medical foundation models has unique challenges and requires domain knowledge since medical data are expensive to collect, annotate interpret, and their quality varies significantly across hospitals and clinical studies.


Real-world images are 2D projections of the 3D world and therefore it is relatively easy to collect a large number of images that cover the variability of object(s)/scenes in terms of viewpoint, angles, scales, appearance and locations. To the contrary, medical images are acquired for a particular clinical purpose, through certain protocols and scanners that require use by an expert who controls the machine settings, including the view angles and scales. The image complexity and variability typically come from scanner differences, scanning protocols, and, most importantly, anatomical and other variations among individuals, disease appearance, location, and stage of the disease.

These unique properties of medical images manifest during the development of public datasets. Classical public datasets were collected for a specific purpose using certain protocols and scanners such as EyePACS~\cite{devente23airogs}, SUN-SEG~\cite{ji2022video}, ISIC~\cite{cassidy2022analysis},Chestx-ray8~\cite{wang2017chestx} CAMELYON~\cite{litjens20181399}, and EndoVis~\cite{allan20202018}, thus usually limited to a single modality and a specific anatomical area for a particular task. Recently, general-purpose datasets have been obtained using multiple protocols and scanners, such as Totalseg~\cite{totalseg}, AMOS~\cite{ji2022amos}, FLARE~\cite{MedIA-FLARE21,FLARE22}, autoPET~\cite{gatidis2022whole}, BraTS~\cite{menze2014multimodal}, ISLES~\cite{hernandez2022isles}, DigestPath~\cite{da2022digestpath}, and MIMICS~\cite{johnson2019mimic}. Correspondingly, the size of datasets has changed from small scale to large scale, and data variability is increasing.
When preparing datasets to train medical foundation models, one should select the most representative cases to train generalizable models and cover corner cases instead of simply collecting large amounts of data. 

In addition, instead of building a unified dataset covering all image modalities (e.g., an``ImageNet''~\cite{deng2009imagenet} type of database for medical data), a more feasible solution is to begin with modality-specific datasets and then attempt to merge those which are complementary.

\noindent\textbf{Data Adaptation for Downstream Tasks}

The emergence of foundation models signals a notable reversal from the previously encouraging trend of model openness and accessibility within the scientific community.
Although pretrained instances of certain models like LLaMA~\cite{touvron2023llama,touvron2023llama2} and SAM~\cite{kirillov2023segany} are publicly accessible,
models such as GPT-3~\cite{brown2020language}, and GPT-4~\cite{openai2023GPT4} are not publicly available; only API access is possible which is restricted to a limited number of users. Furthermore, the datasets employed for training foundation models are not made available to the wider research community. 
The computational and engineering resources required to train foundation models from the ground up are cost prohibitive for academia and most companies. 
This creates a barrier that prevents the vast majority of artificial intelligence researchers from participating in this pivotal ML research and methodology.

Recent advancements in ML have facilitated the efficient adaptation of foundation models for downstream tasks, requiring only a minimal number of training samples (e.g., prompt engineering and efficient methods for retraining). Recent research on the use of ML for medical image analysis has been establishing benchmarks and releasing datasets.~\cite{wang2023medfmc,yi2023towards}. These efforts hold significant promise for enabling the effective deployment of large-scale foundation models to tackle an array of clinical challenges.

\section{Applications and Benefits of Foundation Models}

Leveraging foundation models trained on large datasets to address specific medical needs is crucial for achieving accurate and reliable image analysis and disease diagnosis and prognosis, minimizing the need for data collection, reducing the time and cost associated with data labeling, and upholding patient data privacy and confidentiality.

\noindent\textbf{Long-tailed Problems}

Medical image analysis methods often face the challenging long-tail data scenario, caused by often heavily imbalanced datasets in which many common disease cases coexist with relatively few rare disease cases. Consequently, the scarcity of data for training models to accurately identify these rare cases can lead to significant performance degradation issues.
The few-shot setting aligns perfectly with the long-tailed scenario, which frequently arises in medical imaging when only a few rare disease cases/high-quality annotations are available. 

The initial stage involves selecting the appropriate foundation model (including general, modality-specific, and organ/task-specific models), depending on the application and available resources. Then, data augmentation techniques are employed to augment the few annotated samples and make full use of the available supervised information. 
These techniques include image augmentation methods like rotation, cropping, color transformation, noise injection, and random erasing, as well as image generation techniques, e.g., GAN and diffusion-based models~\cite{chambon2022roentgen,pinaya2022brain,ding2023large}.

Utilizing medical foundation models that have been trained on vast amounts of data can reduce the labeled data required for training, which minimizes the need for manual annotation by medical professionals. Additionally, these models can lead to more reliable diagnoses and treatment decisions.

\noindent\textbf{Explainable and Generalizable Models}


The lack of explainability in deep learning models can lead to distrust issues when clinicians are accustomed to making explainable clinical inferences. Similar to explainability, the generalizability of a model (a model trained on data from one medical center applied to data from other medical centers with significant variations or domain shifts) is also necessary due to the previously mentioned dataset limitations. Therefore, innovative methodologies are needed to improve the explainability and generalizability of the current models in order to be used effectively in clinical practice~\cite{wang2023editorial}.

Foundation models provide a unified framework that can support detection, segmentation, and classification tasks, which is essential for evidence-based decision-making. Moreover, these models are typically trained on large-scale datasets covering a wide range of medical centers, scanners, and protocols, resulting in promising generalizability of the learned feature representations.

\noindent\textbf{Privacy Preserving Methods}

While the computer vision community has an established history of open-sourcing large-scale datasets, such as the ImageNet, making publicly available large amounts of medical data is currently not possible due to regulatory and privacy issues. Foundation models offer an alternative way for knowledge sharing, while protecting patient privacy.
Transfer learning techniques can be used to adapt the foundation model using a smaller dataset of interest, avoiding directly accessing massive raw data.

Furthermore, sharing multi-cohort knowledge in foundation models is also feasible through
the use of federated learning ~\cite{li2020multi,kaissis2021end}, which enables training on data distributed across multiple institutions or devices without the data ever leaving the local machines. This paradigm ensures data privacy when training the foundation models on large distributed datasets. 

Foundation models can also be used to create synthetic data~\cite{ding2023large} which can ensure data privacy preservation. Using generative models to create synthetic medical images which are statistically similar to real medical images, researchers can train models on these synthetic images instead of using real patient data.

\noindent\textbf{Integration with Large Language Models}

The study of vision-language models is gaining prominence due to their capacity for extracting nuanced information and learning superior representations. However, the majority of existing research predominantly concentrates on the analysis of X-ray images in conjunction with their corresponding reports~\cite{zhang2022contrastive,tiu2022expert,zhou2022generalized,lee2023unified}. 
The recent advances in large language models (LLMs) which are trained on vast amounts of text data have significantly improved natural language processing capabilities~\cite{brown2020language,ouyang2022training,touvron2023llama,touvron2023llama2,chowdhery2022palm,google2023palm2}.
The applications are further expanded beyond textual domains with the integration of vision models~\cite{openai2023GPT4,wu2023visual,alayrac2022flamingo,li2023blip,driess2023palme,2023visionllm,liu2023visual}.
By combining language and vision data, these large-scale vision-language models have unlocked exciting possibilities for the future use of medical foundation models.

Integrating medical image analysis systems with general domain LLMs or medical domain LLMs~\cite{singhal2023large,singhal2023towards} holds immense potential for healthcare applications. For instance, these models can be trained to generate descriptive captions for medical images, facilitating automated radiology reports or succinct summaries of complex visuals. Furthermore, decision support systems can also benefit from associating visual features from medical images with text from patient records, providing accurate disease diagnosis and prognosis~\cite{li2023llava,zhang2023large,wang2023chatcad}.
However, these frameworks are still preliminary, as they usually integrate existing LLMs as a module by prompting, without fine-tuning and/or consolidating data modalities such as medical images to these models. The efforts to open-source foundation models and the ability to fine-tune them will be essential in healthcare.

\begin{figure}[h]
  \centering
  \includegraphics[width=0.65\linewidth]{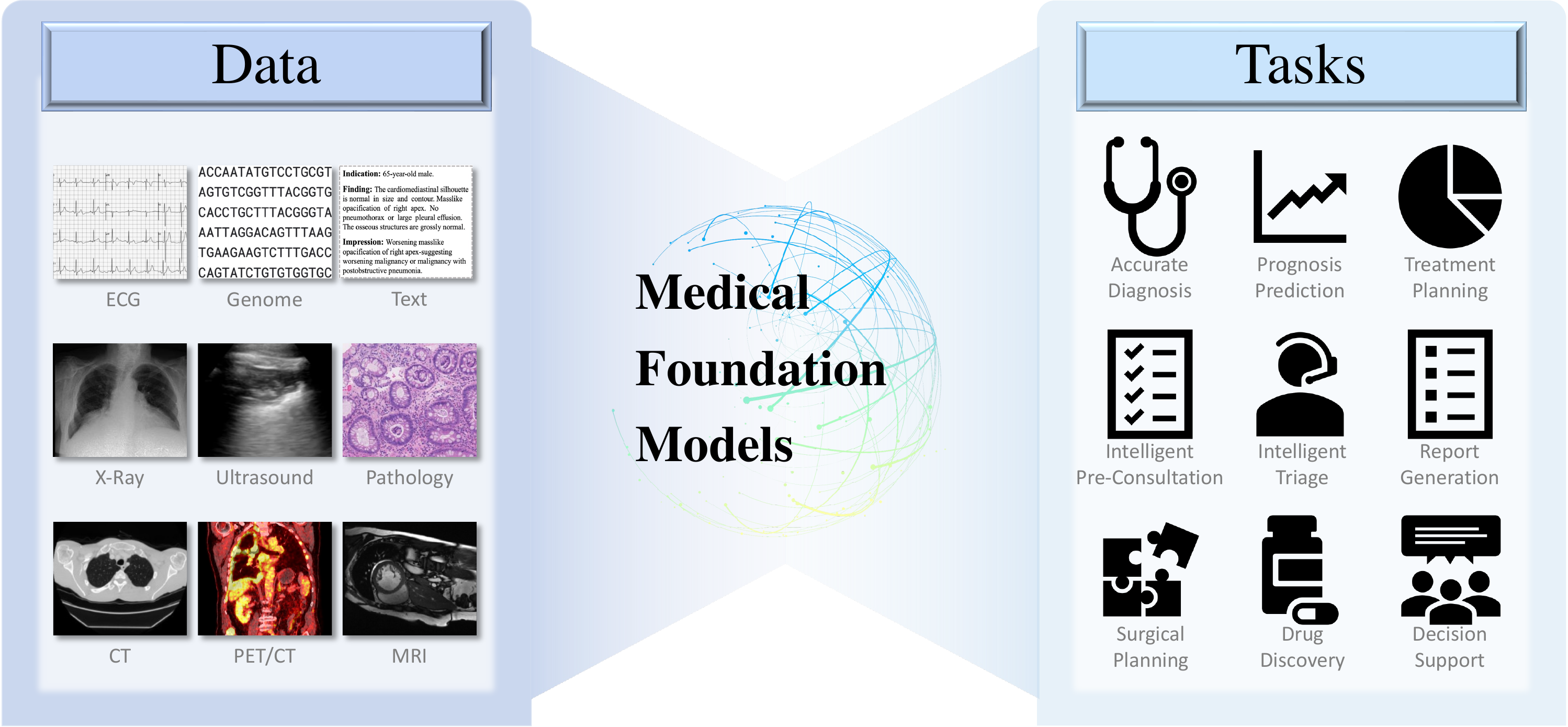}
  \caption{Models trained on multiple medical data modalities can enable comprehensive clinical solutions. }
  \label{fig:app}
\end{figure}

\section{Future Directions of Medical Foundation Models}

We have discussed the challenges and opportunities of foundation models for medical image analysis. These insights can help us design more effective and generalizable foundation models.
Since foundation models have only just begun to transform the way medical image systems are built and deployed worldwide, issues and challenges are still difficult to predict. We advocate that researchers from different institutions and disciplines need to collaborate to investigate and explore the spectrum of foundation models for medical image analysis, and contribute to the open-source community by releasing a family of pretrained models which will work on various imaging modalities.

Future directions include multi-modality foundation models, combining various data types (text, image, video, database, molecule) and scales (molecule, gene, cell, tissue, patient, population). Foundation models hold enormous promise as they can assimilate data from various imaging modalities and can incorporate non-imaging modalities to provide a more comprehensive understanding of a patient's condition and its assessment. By leveraging multi-modality foundation models, medical professionals can achieve a more accurate disease diagnosis and develop personalized treatment plans and disease prognosis. These models can potentially improve the overall quality of medical care by facilitating data sharing among different institutions, leading to more efficient and effective patient care and healthcare.

Advances in multi-modality foundation models can contribute to the development of clinical use cases targeting patients with different background and different diseases. A straightforward application is to support radiologists throughout their workflow, such as drafting structured radiology reports automatically and describing possible abnormalities, disease diagnosis and prognosis, as well as proposed treatment. Another possible use case is to assist surgeons. Integrating image, language, and audio modalities, surgeons can communicate with models to make real-time decisions in the operating room by detecting and identifying the anatomical location of important target structures often not clearly visible.
The quest for comprehensive solutions to these and other medical problems is expected to intensify and the near future and produce the desired results (Fig.~\ref{fig:app}).

\bibliographystyle{unsrtnat}
\bibliography{refs}

\end{document}